\documentclass[aps,showpacs,twocolumn]{revtex4}
\usepackage{amsfonts}
\usepackage{amsmath}
\usepackage{amssymb}
\usepackage{graphicx}

\begin{document}

\title{Spin filter of electrons through a zeeman splitting single quantum dot}
\author{Wenxi Lai}\email{wlai@bistu.edu.cn}
\affiliation{Beijing Information Science and Technology University, Beijing 100192, China}

\author{Mengmeng Zhang}
\affiliation{University of Sciences and Technology Beijing, Beijing 100083, China}

\begin{abstract}
Electron spin filter induced by Zeeman splitting in a few-electron quantum dot coupled to two normal electrodes is studied considering Coulomb blockade effect. Based on the Anderson model and Liouville-von Neumann equation, equation of motion of the system is derived and analytical solutions are achieved. Transport windows for perfectly polarized current, partially polarized current and non-polarized current induced by the Zeeman splitting energy and Coulomb blockade potential are exploited. We will give the relations of voltage, magnetic field and temperature for high quality spin filtering.


\end{abstract}


\maketitle

\begin{center}
\textbf{1. Introduction}
\end{center}

Spin filtering in solid state heterostructure was realized early with external magnetic field~\cite{Esaki} and zero applied field~\cite{Hao,Moodera} based on spin dependent barriers. It has attracted quite attention recently due to its potential applications in quantum computation~\cite{DiVincenzo:1999}, spin measurement~\cite{Deshmukh} and quantum state controlling~\cite{Koppens}. The newly developed spin filters have exploited spin dependent electronic state near the Fermi energy of graphite-metal interface~\cite{Karpan}, angular momentum transfer by optical pumping~\cite{Batelaan}, single spin level in quantum dot (QD)~\cite{Recher,Hanson,Yamagishi}, sandwiching organic-ferromagnet between metals~\cite{Hu}.

In this paper, we study electron spin filer in a electrically defined QD which is coupled to two normal electron leads. We consider Coulomb interaction between two electrons, which allows the QD hold one or two electrons at the most. As a results, there are two types of transport process depending on gate voltage, $0\leftrightarrow1$ type in which electron number fluctuating between $0$ and $1$, and $1\leftrightarrow2$ type in which the QD occupation is changing between $1$ and $2$. External magnetic field splits degenerated electron energy level into two brunches with up spin and down spin, respectively. When the two levels for the different spin polarization are distributed two sides of the Fermi energy level, spin dependent tunneling occurs. This is the main mechanism of spin filter in QD and, in fact, involved in many heterostructure based filters~\cite{Esaki,Hao,Moodera,Karpan}. QD has advantages that few energy levels for transport are naturally available due to highly confinement potential and electronic states can be easily controlled with electrical and optical methods. In addition, based on the Pauli's exclusion principle and spin conservation, spin blockade effect~\cite{Koppens,Shaji} can be achieved by adjusting voltages on the QD. It leads to spin filtering of single electrons with high polarization efficiency.

The spin polarized tunneling in QDs has been studied in previous both theoretically and experimentally~\cite{Recher,Hanson,Yamagishi}. However, realizing a high quality spin filter device is still a challenge for wide practical applications~\cite{Moodera:2007}. Considering the importance of spin polarized source in spintronics, it is necessary to further understand the spin filter effect, especially the energy structure of the systems for creating different spin polarized electrons. In this work, we derive a quantum master equation for describing electron transport in a few-electron QD considering the Coulomb blockade effect. Analytical solutions for the density matrix equations are obtained which is significant for study of the spin dependent tunneling. It is worth to note that master equation which derived from the microscopic theory is an efficient approach to describe electron transport in QDs with the Coulomb blockade effect~\cite{Gurvitz}. We will show here that, in the magnetic field induced spin filter, Zeeman energy is critical important. When the Zeeman splitting is smaller than the bias voltage exerted on the QD, the energy width of spin filtering window is determined by the Zeeman energy. On the other hand, when the Zeeman energy is larger than the bias voltage, the filtering window width would be determined by the bias voltage. Coulomb blockade potential induces transport windows for partial spin polarization. Finally, relations between the Zeeman splitting and temperature for different spin filtering efficiencies are discussed.

\begin{center}
\textbf{2. Model and Equations}
\end{center}

\begin{figure}
\includegraphics[width=8cm]{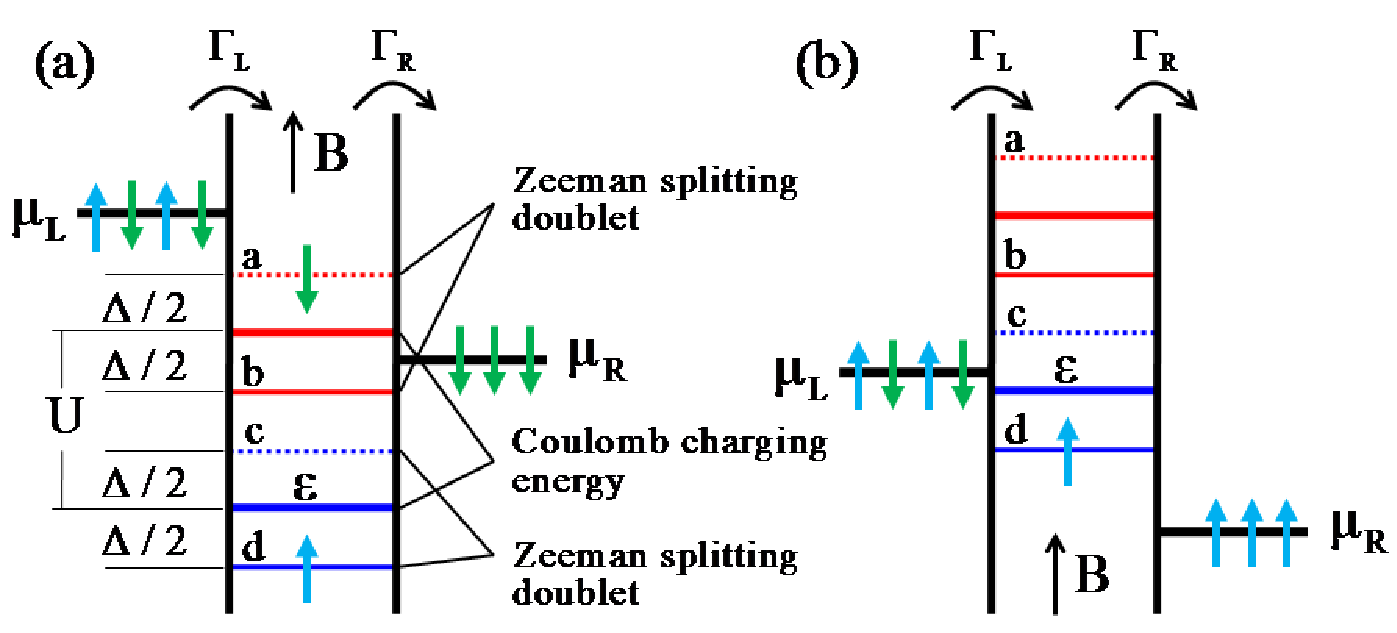}\\
\caption{(Color on line) Schematic illustration of the filter model. (a) Filter with $1\leftrightarrow2$ type electron fluctuation in the QD. (b) Filter with $0\leftrightarrow1$ type electron fluctuation in the QD. Narrow black arrow indicates magnetic field direction. The thick arrows represent electron with definite spin direction. The dotted lines are energy levels for spin down electrons, spin up electrons occupies the solid lines which is separated from the dotted line by $\Delta$. $a, b$ denote the levels (red lines) of Zeeman splitting doublet near the Coulomb potential $\varepsilon+U$, $c, d$ indicate the levels (blue lines) of Zeeman splitting doublet corresponding to the QD level $\varepsilon$.}
\label{sys}
\end{figure}

The typical QD spin filter models are shown in Fig.~\ref{sys}. $\varepsilon$ is electron orbital ground state energy. The energy is split into two levels with distance of $\Delta=g^{*} \mu_{B}B$ under a homogeneous magnetic field $(0,0,B)$ in $z$ direction. $g^{*}$ is the electron $g$ factor in the QD material and $\mu_{B}$ is the Bohr magneton. When the ground state is occupied by one electron, the sequential electron with opposite spin direction needs energy $U+\Delta$ to enter the dot due to Coulomb blockade effect (with charging energy $U$) and Zeeman energy. The QD connected to two leads with left lead characterized by the chemical potential $\mu_{L}$ and right lead characterized by $\mu_{R}$. The Anderson Hamiltonian for the system reads ($\hbar=1$)
\begin{eqnarray}
H &=&\varepsilon c_{\sigma }^{\dag }c_{\sigma }+U c_{\uparrow }^{\dag
}c_{\uparrow }c_{\downarrow }^{\dag }c_{\downarrow }+\sum_{k,\sigma,\alpha
}\epsilon _{\alpha k}c_{\alpha k\sigma }^{\dag }c_{\alpha k\sigma } \notag \\
&&+g^{*} \mu _{B}BS_{z}+ \sum_{\sigma ,k,\alpha}(t_{\alpha
}c_{\alpha k\sigma }^{\dag }c_{\sigma }+t_{\alpha }^{\ast }c_{\sigma }^{\dag
}c_{\alpha k\sigma })
\label{eq:system-Hamiltonian}
\end{eqnarray}
where $c_{\sigma }$ and $c_{\alpha k\sigma }$ are electron annihilation operators with wave vector $k$ and spin state $\sigma=\uparrow,\downarrow$ in the QD and in the leads $\alpha =L,R$. $S_{z}$ is electron spin operator in the $z$ direction of the magnetic field. The last term describes tunneling process between QD and leads with amplitude $t_{\alpha}$ taken as independent on wave vector $k$.
In interaction picture, the system Hamiltonian reads
\begin{eqnarray}
\widetilde{H}_{1}(t)= \sum_{\sigma ,k,\alpha}(t_{\alpha }c_{\alpha
k\sigma }^{\dag }c_{\sigma }e^{i(\epsilon _{\alpha k}-\varepsilon _{\sigma
}-U c_{\bar{\sigma}}^{\dag }c_{\bar{\sigma}})t}+H.c.),
\label{eq:interaction-Hamiltonian}
\end{eqnarray}

which is obtained according to $\widetilde{H}_{1}=e^{iH_{0}t}H_{1}e^{-iH_{0}t}$ with the free propagation Hamiltonian $H_{0}=\sum_{\sigma }\varepsilon _{\sigma }c_{\sigma }^{\dag }c_{\sigma
}+U c_{\uparrow }^{\dag }c_{\uparrow }c_{\downarrow }^{\dag }c_{\downarrow
}+\sum_{k,\sigma,\alpha}\epsilon _{\alpha k}c_{\alpha k\sigma }^{\dag
}c_{\alpha k\sigma }$. Here, the electron energy in quantum dot is written as $\varepsilon _{\uparrow } =\varepsilon +\Delta/2$ and $\varepsilon _{\downarrow } =\varepsilon -\Delta/2$ considering the relation $S_{z}=(c^{\dag}_{\uparrow}c_{\uparrow}-c^{\dag}_{\downarrow}c_{\downarrow})/2$. The spin index $\bar{\sigma}$ in Eq. \eqref{eq:interaction-Hamiltonian} is defined as if $\sigma=\uparrow(\downarrow)$, then $\bar{\sigma}=\downarrow(\uparrow)$.

The whole system satisfies Liouville-von Neumann equation in interaction picture,
\begin{eqnarray}
\frac{\partial \widetilde{\rho }_{tot}(t)}{\partial t}=-i[\widetilde{H}_{1}(t),\widetilde{\rho }_{tot}(t)],
\end{eqnarray}
where $\widetilde{\rho }_{tot}(t)=e^{-iH_{0}t}\rho_{tot}(t)e^{iH_{0}t}$ is the total density matrix operator. The electronic leads are much large compared to the QD, and we assume the two leads are in statistical equilibrium state with time independent equilibrium density matrix $\rho_{leads}$. In this case, the total density matrix can be written in the factorized form $\rho_{tot}(t)=\rho(t)\rho_{leads}$, which is called Born approximation. $\rho(t)$ is reduced density matrix of the QD.

According to the standard derivation of quantum master equation in Markov approximation~\cite{Walls,Scully}, we reach the equation for the QD density matrix $\rho$,
\begin{eqnarray}
\frac{\partial }{\partial t}\rho &=&-i[\sum_{\sigma
}\varepsilon _{\sigma }c_{\sigma }^{\dag }c_{\sigma }+U c_{\uparrow }^{\dag
}c_{\uparrow }c_{\downarrow }^{\dag }c_{\downarrow },\rho ] + \mathcal{L}\rho,
\label{eq:master-equation}
\end{eqnarray}
where
\begin{widetext}
\begin{eqnarray}
\mathcal{L}\rho &=& \frac{1}{2}\sum_{\sigma,\alpha}\Gamma _{\alpha }[f_{\alpha \sigma }(n_{\bar{\sigma}})(c_{\sigma }^{\dag }\rho c_{\sigma }-c_{\sigma }c_{\sigma
}^{\dag }\rho )+(1-f_{\alpha \sigma }(n_{\bar{\sigma}}))(c_{\sigma }\rho c_{\sigma}^{\dag }-c_{\sigma }^{\dag }c_{\sigma }\rho )+H.c.],
\end{eqnarray}
\end{widetext}

In this equation, the tunneling rates are expressed as  $\Gamma _{\alpha }=2 \pi \hbar \left\vert t_{\alpha }\right\vert ^{2}N_{\alpha }(\epsilon )$ with density of states $N_{\alpha }(\epsilon )$ of electrons at energy $\epsilon$. The Fermi distribution function
\begin{eqnarray}
f_{\alpha \sigma }(n_{\bar{\sigma}})=\frac{1}{e^{(\varepsilon _{\sigma
}+U n_{\bar{\sigma}}-\mu _{\alpha})/k_{B}T}+1}
\label{eq:Fermi-distrib}
\end{eqnarray}
depends on the charging energy $U$ conditioned by the occupation, $n_{\bar{\sigma}}=c_{\bar{\sigma}}^{\dag }c_{\bar{\sigma} }$, of electron with opposite spin direction. $k_{B}$ is Boltzmann constant and $T$ is electron temperature.

The electron current on the left side $I_{L}$ and right side $I_{L}$ of the QD is related to the charge fluctuation in the QD as
\begin{eqnarray}
e\frac{\partial n}{\partial t}=I_{L}-I_{R}.
\label{eq:charge-fluct}
\end{eqnarray}
where $n=\sum_{\sigma}c_{\sigma }^{\dag }c_{\sigma }$ and $e$ is absolute value of one electron charge. We concentrate on the current detected on the right side of the QD, by denoting $I_{R}$ with $I$ for convenience, and the current can be derived based on Eqs. \eqref{eq:master-equation} and \eqref{eq:charge-fluct} as
\begin{widetext}
\begin{eqnarray}
I_{\uparrow} &=&-e\Gamma _{R}[f_{R \uparrow}(0)\rho _{00}+f_{R \uparrow}(1)\rho
_{22}-(1-f_{R \uparrow}(0))\rho _{11}-(1-f_{R \uparrow}(1))\rho _{33}]
\label{eq:up-currt}
\end{eqnarray}
for spin up electron and
\begin{eqnarray}
I_{\downarrow} &=&-e\Gamma _{R}[f_{R \downarrow}(0)\rho _{00}+f_{R \downarrow}(1)\rho _{11}-(1-f_{R \downarrow}(0))\rho _{22}-(1-f_{R \downarrow}(1))\rho_{33}]
\label{eq:down-currt}
\end{eqnarray}
\end{widetext}
for spin down electron. The Fermi distribution functions in the current formulas $I_{\uparrow}$ and $I_{\downarrow}$ are given by Eq.\eqref{eq:Fermi-distrib}. Here, $n_{\bar{\sigma}}$ takes its eigenvalue $0$ or $1$.

The up polarized and down polarized electron current can be calculated using Eqs.\eqref{eq:up-currt}, \eqref{eq:down-currt}, and combining with the solution of Eq.\eqref{eq:master-equation}. To solve the density matrix elements, we project Eq.\eqref{eq:up-currt} into the Hilbert space that generated from the QD eigenstates. The QD all together has $4$ states, they are empty state $|0\rangle$, one electron occupied state with spin up  $|\uparrow\rangle$, one electron occupied state with spin down  $|\downarrow\rangle$ and two electron occupied state with opposite directed spins $|\uparrow \downarrow\rangle$. For the convenience of description, we encode these state by $|0\rangle=|0\rangle$, $|1\rangle=|\uparrow\rangle$, $|2\rangle=|\downarrow\rangle$ and $|3\rangle=|\uparrow \downarrow\rangle$. In the condition of stationary state, $\frac{\partial\rho}{\partial t}=0$, we obtain general solutions for Eq. \eqref{eq:master-equation} as follows:
\begin{eqnarray}
\rho _{00}=\frac{A}{A+B+C+D},
\label{eq:R00}
\end{eqnarray}
\begin{eqnarray}
\rho _{11}=\frac{B}{A+B+C+D},
\label{eq:R11}
\end{eqnarray}
\begin{eqnarray}
\rho _{22}=\frac{C}{A+B+C+D},
\label{eq:R22}
\end{eqnarray}
and
\begin{eqnarray}
\rho _{33}=\frac{D}{A+B+C+D},
\label{eq:R33}
\end{eqnarray}

where $A$, $B$, $C$ and $D$ consist of coefficients of Eq.\eqref{eq:master-equation} (see Appendix). The diagonal density matrix elements $\rho _{00}$, $\rho _{11}$, $\rho _{22}$ and $\rho _{33}$ represent distribution probabilities of the QD in its $4$ eigenstates respectively. Off diagonal elements of the density matrix $\rho$ are not coupled to these diagonal terms.

\begin{figure}
  \includegraphics[width=9cm]{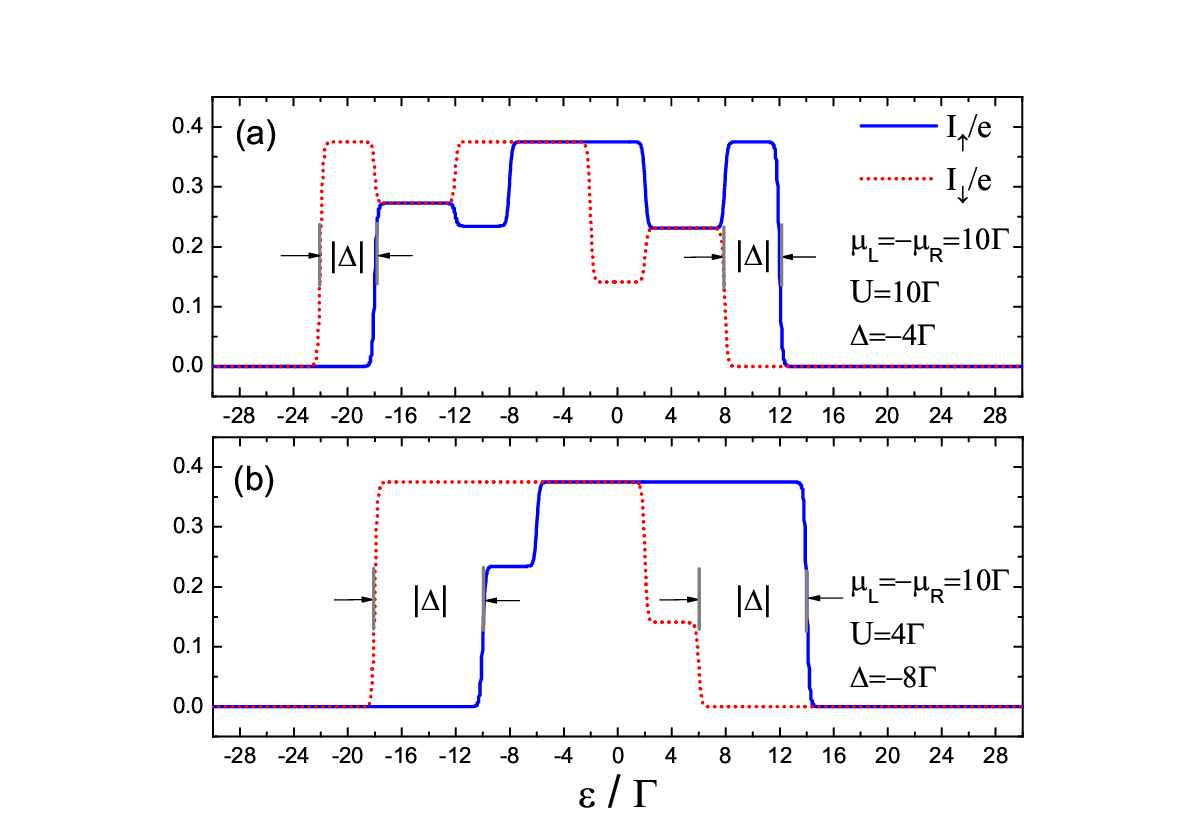}\\
  \caption{(Color on line) Current as a function of the gate voltage $\varepsilon$. (a) In the condition $|\mu_{L}-\mu_{R}| > U > |\Delta|$. (b) In the condition $|\mu_{L}-\mu_{R}| > |\Delta| > U$. The rest parameters are $\Gamma_{R}=0.6\Gamma$, $k_{B}T=0.1\Gamma$}
  \label{voltage}
\end{figure}

\begin{center}
\textbf{4. Results}
\end{center}

\subsection{Perfect spin filter at low temperature}

In this subsection, we consider low temperature limit in which the thermal energy much smaller than other energy scales, i.e., $|\mu_{L}-\mu_{R}|, U, \Delta, \Gamma_{L}, \Gamma_{R} \gg k_{B}T$. It helps us to more clearly discuss the relations between other energies. We take the left tunneling rate $\Gamma_{L}$ as the unit energy by writing $\Gamma_{L}=\Gamma$. Since, $g^{*}$ is always negative for electron, we have $\Delta<0$ for the magnetic field in $z$ direction. It leads to spin up (down) electron occupies lower (higher) energy level as shown in Fig.~\ref{sys}. In contrast, for magnetic field in the $-z$ direction, spin up (down) electron occupies higher (lower) energy level, which is not considered in the present work.

Our main purpose is to systematically find the condition of maximal spin polarized current. To this end, we classify the rest parameters, bias voltage $\mu_{L}-\mu_{R}$, Zeeman splitting $\Delta=g^{*} \mu_{B}B$ and Coulomb potential $U$. First, we take the bias voltage is larger than the other two energy scales, $|\mu_{L}-\mu_{R}| > U, |\Delta|$ as shown in Fig.\ref{voltage}.

Now, for the energy configuration $|\mu_{L}-\mu_{R}|> U > |\Delta|$ in Fig. \ref{voltage}(a), currents versus dot level $\varepsilon$ appears interesting spectrum. The currents for spin up and spin down electron manifest different behaviors. There are all together $5$ kinds of current distributions regime with respect to the QD level $\varepsilon$. As detail discussion, we concentrate on the QD level range $-28\Gamma<\varepsilon<-5\Gamma$, which contains all these $5$ kinds of current behaviors. (1) For dot level $\varepsilon<-22\Gamma$, no current can be observed due to there is no energy level included in the transport window. Using Eqs.\eqref{eq:up-currt} and \eqref{eq:down-currt}, we have the simplified current formulas $I_{\uparrow}(\varepsilon) =-e\Gamma _{R}[\rho _{00}+\rho_{22}]$ and $I_{\downarrow}(\varepsilon) =-e\Gamma _{R}[\rho _{00}+\rho_{11}]$ at low temperature limit. Using Eqs.\eqref{eq:R00}, \eqref{eq:R11}, \eqref{eq:R22} and \eqref{eq:R33}, density matrix elements can be obtained as $\rho_{00}(\varepsilon)=\rho_{11}(\varepsilon)=\rho_{22}(\varepsilon)=0$, $\rho_{33}(\varepsilon)=1$. The result reflects that the QD has only possible state of two-electron occupation with zero current $I_{\uparrow}(\varepsilon)=I_{\downarrow}(\varepsilon)=0$. (2) Within the dot level range $-22\Gamma>\varepsilon>-18\Gamma$, fully polarized filter is obtained with pure down spins. The current values are given by the strict calculation in Fig.\ref{voltage} (a). It can be estimated at low temperature limit. In this case we have $I_{\downarrow}(\varepsilon) =-e\Gamma _{R}[\rho _{00}-\rho_{33}]$ and  $I_{\uparrow}(\varepsilon) =-e\Gamma _{R}[\rho _{00}+\rho_{22}]$, where $\rho _{00}(\varepsilon)=0$,  $\rho _{11}(\varepsilon)=\Gamma_{R}/(\Gamma_{L}+\Gamma_{R})$, $\rho _{22}(\varepsilon)=0$, $\rho_{33}(\varepsilon)=\Gamma_{L}/(\Gamma_{L}+\Gamma_{R})$. It leads to the current for spin up $I_{\uparrow}(\varepsilon)=0$ and for spin down
\begin{eqnarray}
I_{\downarrow}(\varepsilon) =e\frac{\Gamma _{R}\Gamma_{L}}{(\Gamma_{L}+\Gamma_{R})},
\end{eqnarray}
which is the typical current formula of the single level QD tunneling at low temperature~\cite{Lu}. Indeed, Only the highest level $\texttt{a}$ of the QD, which allows the occupation of spin down electron, is involved in the transport (see Fig.\ref{sys}(a)). From the solution of density matrix elements, it can be deduced that QD state is fluctuating between the occupation of single electron with spin up and two electrons with opposite directed spins. It is explained by the fact that an electron with spin up is bounded in the QD and another electron with spin down is hopping through the dot. Since, the energies of these tunneling electrons are above the charging energy, the current appeared to be not affected by the Coulomb blockade interaction. (3) The third kind of current distribution is in the regime of  $-18\Gamma<\varepsilon<-12\Gamma$. According to Eqs. \eqref{eq:up-currt} and \eqref{eq:down-currt}, we have the currents $I_{\downarrow}(\varepsilon)= I_{\uparrow}(\varepsilon)=-e\Gamma _{R}[\rho _{00}-\rho_{33}]$. Our derivations give $\rho_{00}(\varepsilon)=0$, $\rho_{11}(\varepsilon)=\rho_{22}(\varepsilon)=\Gamma _{R}/(\Gamma _{L}+2\Gamma _{R})$ and $\rho_{33}(\varepsilon)=\Gamma _{L}/(\Gamma _{L}+2\Gamma _{R})$. Then the currents read
\begin{eqnarray}
I_{\uparrow}(\varepsilon)=e\frac{\Gamma _{R}\Gamma _{L}}{(\Gamma _{L}+2\Gamma _{R})}
\end{eqnarray}
and $I_{\downarrow}(\varepsilon)= I_{\uparrow}(\varepsilon)$, which is in agreement with the result given by the approach of many-body Schrodinger equation ~\cite{Gurvitz}. In the regime of QD, one level $\texttt{a}$ for spin down and another level $\texttt{b}$ for spin up are involved in the transport. Therefore we observe current with both up and down electron spins. The total current is increased in contrast to that in case (2), however, the pure down polarized current depressed comparing that in the case (2) due to Coulomb interaction between two sequentially tunneling electrons. Since spin up electron is not absolutely bounded now. Instead, it can be excited by the Coulomb interaction and jump onto the level $\texttt{b}$ and transfer to the right lead. Whereas, the occupation of the $\texttt{b}$ level affects an down polarized electron passing through the level $\texttt{a}$ due to the Coulomb interaction between Zeeman doublet $\texttt{a}$ and $\texttt{b}$. At low temperature, since the Zeeman doublet levels are fully included in the transport window, the two electron currents with different spin direction have the same amplitudes. (4) Further moving the QD level upward, we turn to the regime $-12\Gamma<\varepsilon<-8\Gamma$, where three levels $\texttt{a}$, $\texttt{b}$, $\texttt{c}$ are involved in the transport. At low temperature limit, we have the currents $I_{\downarrow}(\varepsilon)= e\Gamma _{R}[\rho _{22}+\rho_{33}]$ and $I_{\uparrow}(\varepsilon)=-e\Gamma _{R}[\rho _{00}-\rho_{33}]$ and the occupation probabilities $\rho_{00}(\varepsilon)=\Gamma _{L}\Gamma _{R}^{2}/2(\Gamma _{R}+\Gamma _{L})^{3}$, $\rho_{11}(\varepsilon)=(2\Gamma _{R}\Gamma _{L}^{2}+3\Gamma _{L}\Gamma _{R}^{2}+2\Gamma_{R}^{3})/2(\Gamma _{R}+\Gamma _{L})^{3}$, $\rho_{22}(\varepsilon)=(2\Gamma _{R}\Gamma _{L}^{2}+\Gamma _{L}\Gamma _{R}^{2})/2(\Gamma _{R}+\Gamma _{L})^{3}$ and $\rho_{33}(\varepsilon)=(2\Gamma _{R}\Gamma _{L}^{2}+\Gamma _{L}\Gamma _{R}^{2}+2\Gamma_{L}^{3})/2(\Gamma _{R}+\Gamma _{L})^{3}$. Then the currents read
\begin{eqnarray}
I_{\downarrow}(\varepsilon)=e\frac{\Gamma _{R}\Gamma _{L}}{(\Gamma _{R}+\Gamma _{L})}
\end{eqnarray}
for spin down electrons and
\begin{eqnarray}
I_{\uparrow}(\varepsilon)=e\frac{\Gamma _{R}\Gamma _{L}^{2}}{(\Gamma _{R}+\Gamma _{L})^{2}}
\end{eqnarray}
for spin up electrons. When an electron with spin up occupies the ground state level $\texttt{d}$, it is unable to directly tunnel to the right lead due to the high barrier. However, the electron-electron Coulomb interaction excites the spin up electron to the higher level $\texttt{b}$, which gives rise to the spin up electron transfer to the right lead. Occupation of the level $\texttt{b}$ by the spin up electron prevents the spin down electron to take the level $\texttt{a}$. This Process is similar to the case in (3). However, now, the spin down electron has another choice that it can take level $\texttt{c}$ to pass the QD. On the other hand, when a spin up electron is in the level $\texttt{d}$, the spin down electron would take the level $\texttt{a}$. In a word, the spin down electron tunneling in this process is not prevented by both the Coulomb blockade and spin blockade effects. Therefore, the electron current with spin down is characterized by the maximum current of single level dot. In contrast, the up spin electron if and only if take the level $\texttt{b}$ (which is excited by the Coulomb interaction) for tunneling. As a result, its current is depressed. (5) QD level in the regime $-8\Gamma<\varepsilon<-5\Gamma$ allows all the four levels $\texttt{a}, \texttt{b}, \texttt{c}, \texttt{d}$  join the transport. At low temperature limit, the analytical solutions offer probabilities of the QD states $\rho_{00}(\varepsilon)=\Gamma _{R}^{2}/(\Gamma _{R}+\Gamma _{L})^{2}$, $\rho_{11}(\varepsilon)=\rho_{22}(\varepsilon)=\Gamma _{R}\Gamma _{L}/(\Gamma _{R}+\Gamma _{L})^{2}$ and $\rho_{33}(\varepsilon)=\Gamma _{L}^{2}/(\Gamma _{R}+\Gamma _{L})^{2}$ and currents $I_{\downarrow}(\varepsilon)=e \Gamma_{R} (\rho_{22}+\rho_{33})$ and $I_{\uparrow}(\varepsilon)=e \Gamma_{R} (\rho_{11}+\rho_{33})$. They give rise to currents in terms of the tunneling rates
\begin{eqnarray}
I_{\uparrow}(\varepsilon)=e \frac{\Gamma_{L}\Gamma_{R}}{(\Gamma_{L}+\Gamma_{R}) }
\end{eqnarray}
and $I_{\downarrow}(\varepsilon)=I_{\uparrow}(\varepsilon)$. In this case, the QD works as two-level system, in which the two levels are independent each other with each level contributes current $e \Gamma_{L}\Gamma_{R}/(\Gamma_{L}+\Gamma_{R})$. Since the joining of these $4$ levels indicates that after the QD is occupied by one electron, the second electron can offer energy higher than the Coulomb blockade charging energy. It works for both spin up and down electrons and, therefore, makes tunneling avoids the affect from Coulomb interaction. Until now, we have discussed all the $5$ kinds of current behaviors, taking the left side lines of the Fig.\ref{voltage} (a). The right side lines in the Fig.\ref{voltage} (a) can be explained in analogous way.

\begin{figure}
\includegraphics[width=9cm]{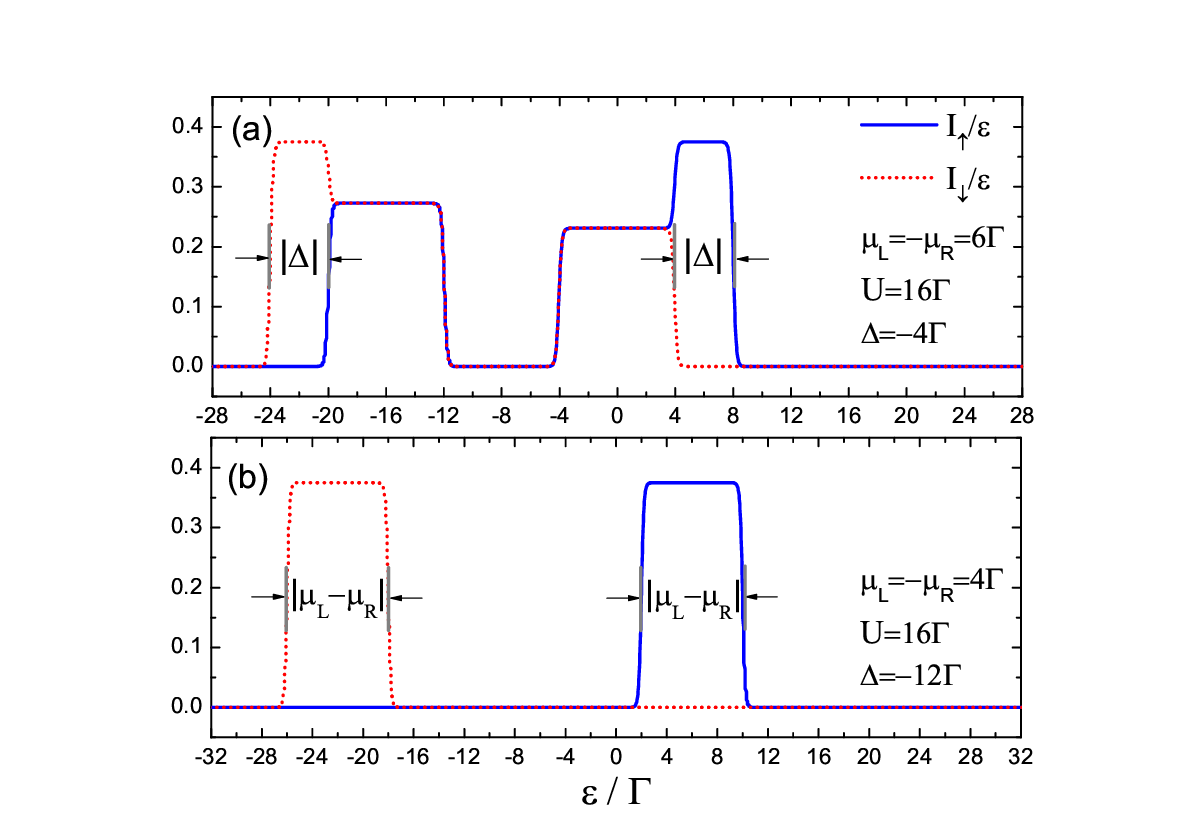}\\
\caption{(Color on line) Current versus the gate voltage $\varepsilon$. (a) In the condition $U > |\mu_{L}-\mu_{R}| > |\Delta|$. (b) In the condition $U > |\Delta| > |\mu_{L}-\mu_{R}|$. The rest parameters are $\Gamma_{R}=0.6\Gamma$, $k_{B}T=0.1\Gamma$}
\label{coulomb}
\end{figure}

\begin{figure}
\includegraphics[width=9cm]{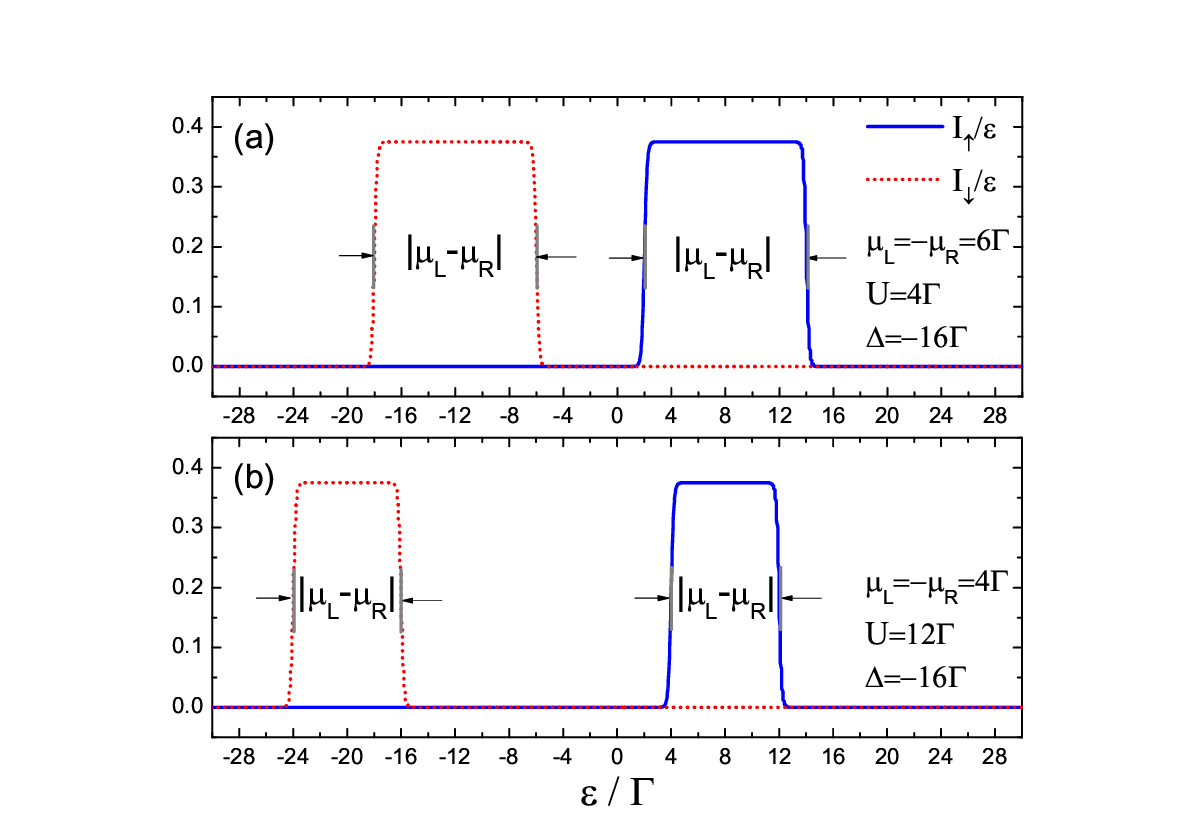}\\
\caption{(Color on line) Current as a function of the gate voltage $\varepsilon$. (a) In the condition $|\Delta| > |\mu_{L}-\mu_{R}|>U $. (b) In the condition $|\Delta| > U > |\mu_{L}-\mu_{R}|$. The rest parameters are $\Gamma_{R}=0.6\Gamma$, $k_{B}T=0.1\Gamma$}
\label{zeeman}
\end{figure}

In Fig.\ref{voltage} (a), we have considered $|\mu_{L}-\mu_{R}|> U > |\Delta|$ and observed that the system have energy level regime of $2 |\Delta|$ for a fully polarized current. By setting $|\mu_{L}-\mu_{R}| > |\Delta|> U$ in Fig.\ref{voltage} (b), we also obtain the QD level range of $2 |\Delta|$ for the fully polarization. The rest four possible energy scale structures are considered, by taking $U > |\mu_{L}-\mu_{R}|> |\Delta|$, $U > |\Delta|> |\mu_{L}-\mu_{R}|$ in Fig. \ref{coulomb} and  $|\Delta|>U > |\mu_{L}-\mu_{R}|$, $|\Delta| > |\mu_{L}-\mu_{R}|> U$ in Fig. \ref{zeeman}. The spectrum characters of electron currents in these figures are included in Fig.\ref{voltage} (a), thereby can be explained in similar way as the above discussion. The most key points that we conclude from these figures can be summarized in a few words. One can achieve absolutely polarized current, partially polarized current and non-polarized current by tuning the gate voltage of a QD. The perfectly polarized current is the pure effect generated from the Zeeman splitting. The partially polarized current is created due to the contribution from both the Zeeman splitting and the Coulomb blockade effect. The non-polarized current is influenced by Coulomb blockade interaction only or not affected by any of these interactions.

\begin{figure}
\includegraphics[width=9cm]{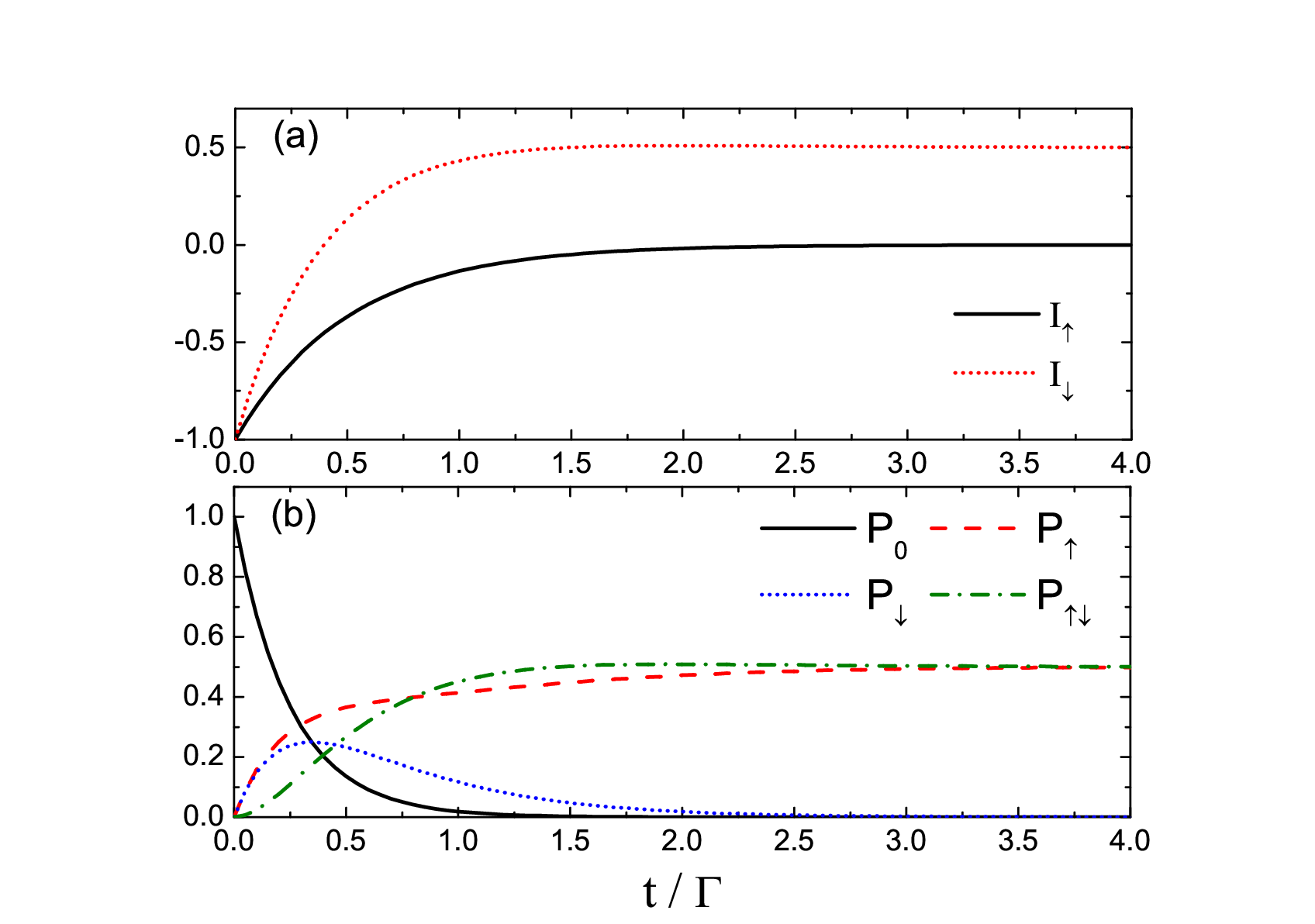}\\
\caption{(Color on line) (a) Current as a function of time $t$ for $\varepsilon=-20\Gamma$. (b) Probabilities of electron occupation with different spin and electron number for $\varepsilon=-20\Gamma$. The rest parameters are $\Gamma_{L}=\Gamma_{R}$, $\mu_{L}=-\mu_{R}=10\Gamma$, $\Delta=4\Gamma$, $U=10\Gamma$ and $k_{B}T=0.1\Gamma$. }
\label{it1}
\end{figure}

\begin{figure}
\includegraphics[width=9cm]{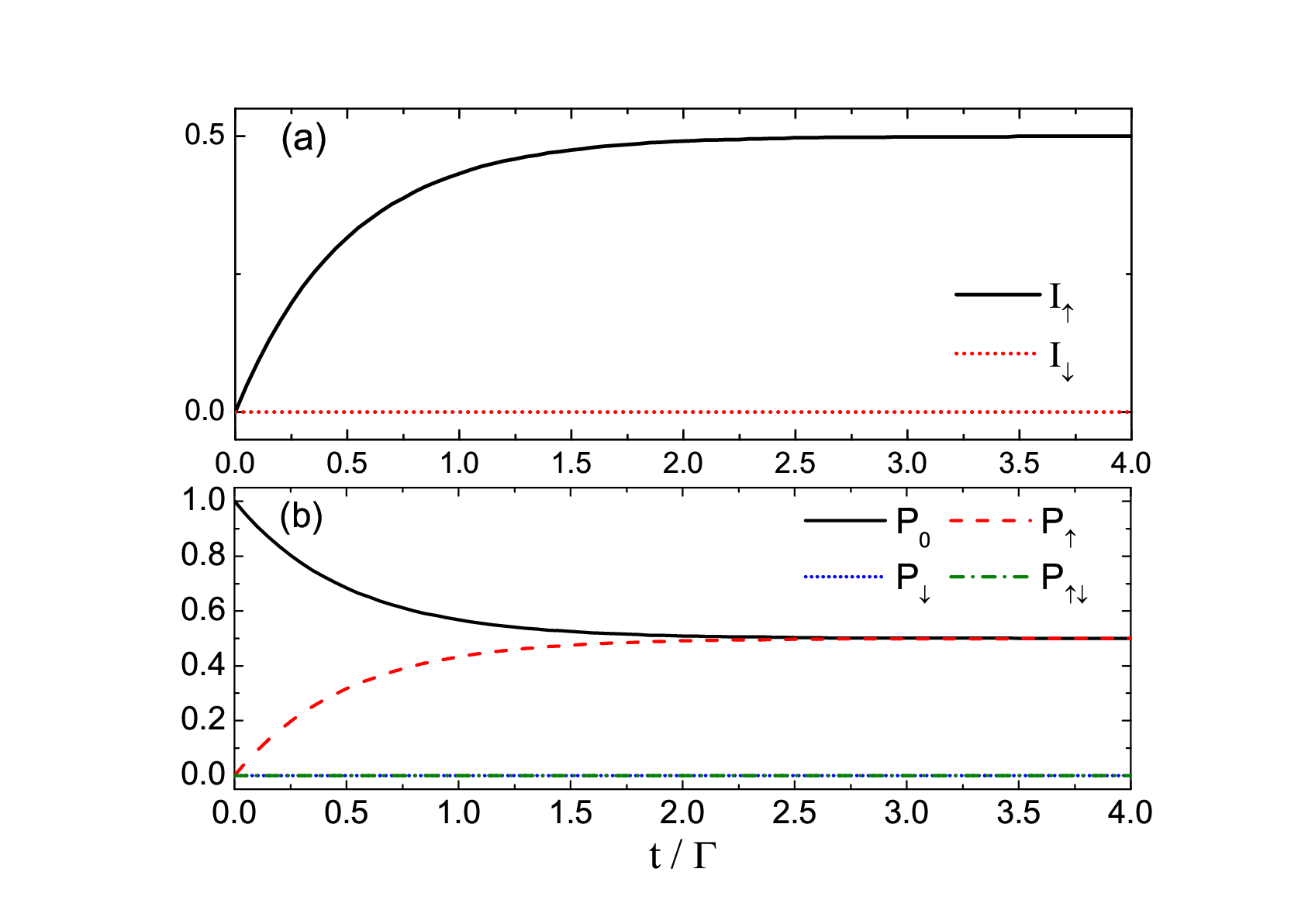}\\
\caption{(Color on line)  (a) Current as a function of time $t$ for $\varepsilon=10\Gamma$. (b) Probabilities of electron occupation with different spin and electron number for $\varepsilon=10\Gamma$. The rest parameters are $\Gamma_{L}=\Gamma_{R}$, $\mu_{L}=-\mu_{R}=10\Gamma$, $\Delta=4\Gamma$, $U=10\Gamma$ and $k_{B}T=0.1\Gamma$. }
\label{it2}
\end{figure}

From Figs. \ref{voltage}, \ref{coulomb}, \ref{zeeman} we further conclude the following results. The fully polarized current has two regimes, one of them is located on the left part of coordinate $\varepsilon$ from $\mu_{R}-U+\Delta/2$ to $\mu_{R}-U-\Delta/2$ for $|\mu_{L}-\mu_{R}| > |\Delta|$ and from $\mu_{R}-U+\Delta/2$ to $\mu_{L}-U+\Delta/2$ for $|\mu_{L}-\mu_{R}| < |\Delta|$. The other regime is on the right part of coordinate $\varepsilon$ from $\mu_{L}+\Delta/2$ to $\mu_{R}-\Delta/2$ for $|\mu_{L}-\mu_{R}| > |\Delta|$ and from $\mu_{R}-\Delta/2$ to $\mu_{L}-\Delta/2$ for $|\mu_{L}-\mu_{R}| < |\Delta|$. Therefore, the QD has a transport window range of $2 \Delta$ for full polarization current when  $|\mu_{L}-\mu_{R}| > |\Delta|$, and has a transport window range of $2 |\mu_{L}-\mu_{R}|$ for the perfect filtering when $ |\Delta|>|\mu_{L}-\mu_{R}|$. In the left side regime, the levels (e.g., \texttt{b} and \texttt{d}) for spin up electron are lower than chemical potentials of both the left and right leads (see Fig. \ref{sys}(a)). It causes to one electron with up polarization be bounded in the lowest level of the dot. In this case, the spin blockade effect allows only spin down electros are allowed to enter the QD and contribute to the current~\cite{Koppens}. The time evolution of the system in Fig.\ref{it1} reflects this spin filter effect with $1\leftrightarrow2$ type electron fluctuation in the QD. The probabilities distributed over the $4$ intrinsic QD eigenstates are calculated through $P_{0}=\rho_{00}$, $P_{\uparrow}=\rho_{11}$, $P_{\downarrow}=\rho_{22}$, $P_{\uparrow\downarrow}=\rho_{33}$. In the right side full polarization regime, the levels (e.g., \texttt{a} and \texttt{c}) for spin down electron is higher than chemical potentials of both the left and right leads (see Fig. \ref{sys}(b)). In this case down electron has not enough energy to enter the QD and only up spin electron transfer the QD. It is the perfectly up spin polarized spin filter, which coincides with the experiment~\cite{Yamagishi}. Such spin filtering of $0\leftrightarrow1$ type electron fluctuation in QD is shown in Fig.\ref{it2}.

\subsection{Temperature dependent polarization efficiency}

Most of the solid state spin filters work based on the fact that Fermi level located between the two energy levels for different spin states~\cite{Esaki,Hao,Moodera,Karpan}. That means the Fermi level is required to be narrow enough to identify the two pin dependent levels, e.g. the Zeeman splitting doublet. As a result, temperature which broadens the Fermi level becomes one of the main factors to weaken the spin filter efficiency. Therefore, to increase the filtering efficiency, one approach is cooling the system, and another method is strengthening the effect magnetic field.

\begin{figure}
\includegraphics[width=9cm]{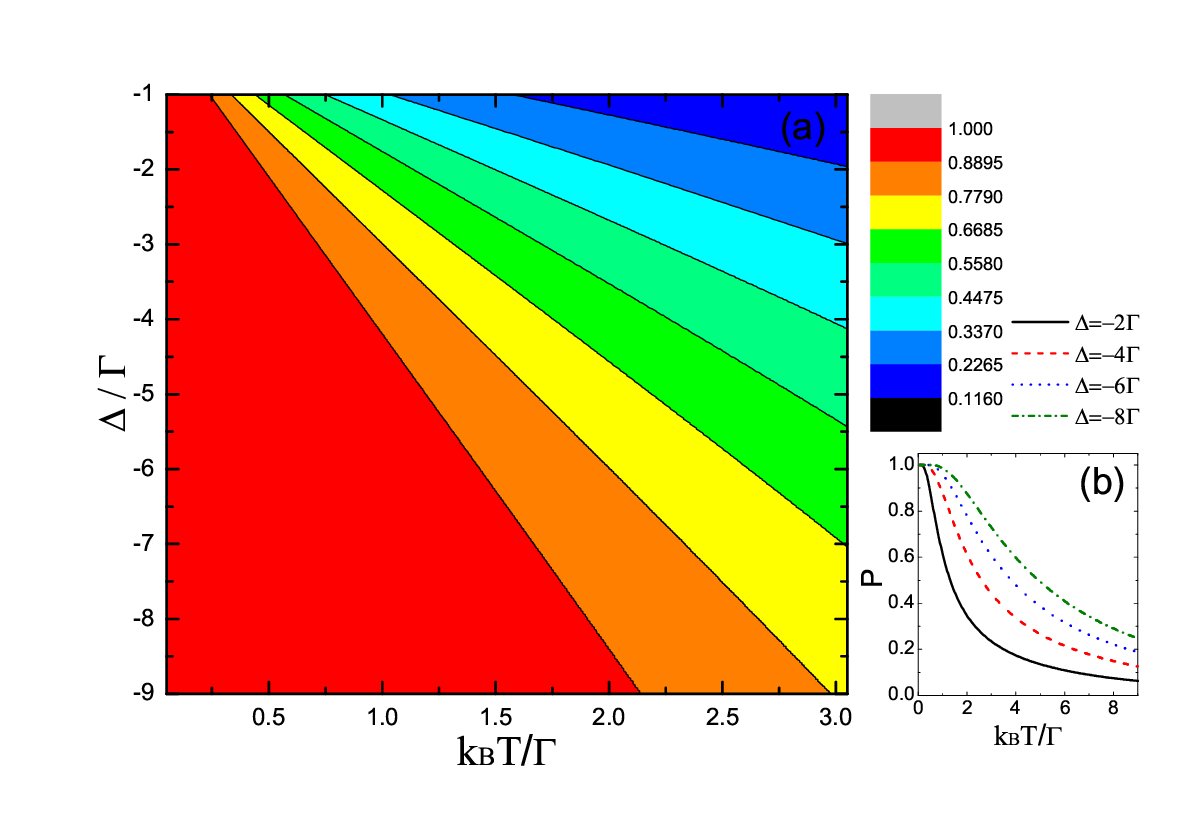}\\
\caption{(Color on line) (a) Polarization efficiency as a function of the Zeeman splitting $\Delta$ and temperature $T$ for the condition $|\mu_{L}-\mu_{R}| > U > |\Delta|$. (b) Polarization efficiency versus temperature. The corresponding parameters are $\varepsilon=10\Gamma$, $\Gamma_{R}=0.6\Gamma$, $\mu_{L}=-\mu_{R}=10\Gamma$, $U=10\Gamma$. }
\label{tbp1}
\end{figure}

\begin{figure}
\includegraphics[width=9cm]{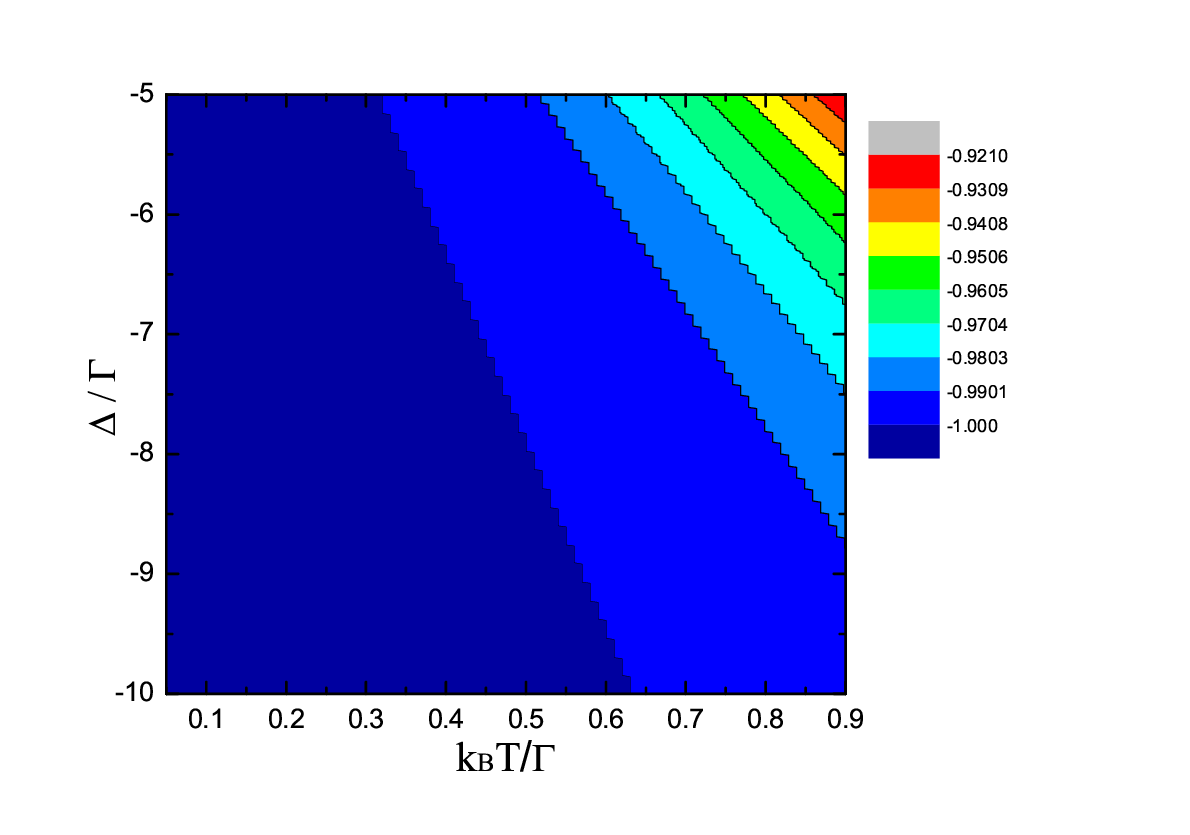}\\
\caption{(Color on line) Polarization efficiency as a function of the Zeeman splitting $\Delta$ and temperature $T$ for the condition $U > |\mu_{L}-\mu_{R}| > |\Delta|$. The corresponding parameters are $\varepsilon=-22\Gamma$, $\Gamma_{R}=0.6\Gamma$, $\mu_{L}=-\mu_{R}=6\Gamma$, $U=16\Gamma$.  }
\label{tbp2}
\end{figure}

Let us see how temperature influences the polarization efficiency for different Zeeman splitting. We used the formula of current polarization  $P=(I_{\uparrow}-I_{\downarrow})/(I_{\uparrow}+I_{\downarrow})$ which takes the value range $-1\leq P \leq1$. Fig.~\ref{tbp1}(a) shows polarization efficiency of the filtering window that introduced in Fig.~\ref{voltage} (a) at the energy center $\varepsilon=10\Gamma$. The polarization is positive which indicates the current is spin up polarized. We find linear relation between Zeeman splitting (or magnetic field) and the thermal energy (or temperature) for any given polarization efficiency. For the thermal energy several time larger than the tunneling rate, current polarization is remarkably destroyed, and the spin up current becomes comparable to the spin down current (see Fig.~\ref{tbp1} (b)). The Polarization efficiency for down spin current is shown in Fig.~\ref{tbp2}. The current is calculated from the filtering window in Fig.~\ref{coulomb} (a) at energy center $\varepsilon=-22$. It is easily observed in this figure that current of very pure spins can be obtained with efficiency higher then $99$ percent for $\Delta > 10 k_{B}T$. As an example, we used the parameters for electrically defined GaAs/AlGaAs QD in Ref.~\cite{Petta}. The electron g factor in GaAs is $g^{*}=-0.44$. At the low temperature $T=135 mK$, thermal energy is $k_{B}T=11.6 \mu e V$. Therefore, for this system, to achieve a spin filter with efficiency higher than $99$ percent, magnetic field strength is required to be larger than $4.5T$.

\begin{center}
\textbf{4. Conclusions}
\end{center}

A QD level is split by a static magnetic field and these magnetically created levels are spin dependent. Choosing one of the pin dependent QD levels in the transport window, a spin filter effect can be obtained. A Markovian quantum master equation describing the spin filtering model is derived considering Coulomb blockade effect. With a finite Zeeman splitting at low temperature, there are two fully polarized spin filter regimes, one is for spin up polarization and another is for spin down polarization. The two regimes can be found by tuning the QD level with gate voltage. Of course, equivalently, one can observe filtering phenomenon by relatively adjusting bias voltage of electrodes instead of the gate voltage. For bias voltage potential larger than the Zeeman splitting energy, $|\mu_{L}-\mu_{R}|>\Delta$, each spin filtering window has a width of the Zeeman energy $\Delta$. The opposite condition that bias voltage potential is smaller than the Zeeman splitting energy, $|\mu_{L}-\mu_{R}|<\Delta$, yields each filtering window width to be equal to the bias voltage potential $|\mu_{L}-\mu_{R}|$. Coulomb blockade effect does not substantially change the filtering effect. Whereas, it induces new transport windows for partially polarized spin filter. The Zeeman splitting spin filter remarkably sensitive to temperature. For definite polarization, temperature and magnetic field satisfy a linear relation. It means if temperature increases, one have to strengthen the magnetic field almost the same proportion as the temperature increase to recover the original filtering efficiency.

\begin{acknowledgments}
This work was supported by the 2019 Research Foundation of Beijing Information Science and Technology University under Grant No. 1925029.
\end{acknowledgments}

\begin{center}
\textbf{Appendix: Expressions of $A$, $B$, $C$ and $D$}
\end{center}
In the general expressions of density matrix elements $\rho _{00}$, $\rho _{11}$, $\rho _{22}$ and $\rho _{33}$ given in  Eq.\eqref{eq:R00}-\eqref{eq:R33}, $A$, $B$, $C$ and $D$ may be written as
\begin{eqnarray}
A=-T_{12}T_{24}T_{33}-T_{13}T_{22}T_{34},
\end{eqnarray}
\begin{eqnarray}
B=T_{11}T_{24}T_{33}+T_{13}T_{21}T_{34}-T_{13}R_{24}T_{31},
\end{eqnarray}
\begin{eqnarray}
C=T_{12}T_{24}T_{31}-T_{12}T_{21}T_{34}+T_{11}T_{22}T_{34},
\end{eqnarray}
and
\begin{eqnarray}
D=T_{13}T_{22}T_{31}+T_{12}T_{21}T_{33}-T_{11}T_{22}T_{33}.
\end{eqnarray}

In the above equations, $T_{i,j}$ ($i,j=1,2,3,4$) are coefficients of the differential equations which are obtained from Eq.\eqref{eq:master-equation} in the system Hilbert space. These coefficients can be expressed as follows:  $T_{11}=-\Gamma_{L}(f_{L}(\varepsilon_{\uparrow})+f_{L}(\varepsilon_{\downarrow}))-\Gamma_{R}(f_{R}(\varepsilon_{\uparrow})+f_{R}(\varepsilon_{\downarrow}))$,
$T_{12}=\Gamma_{L}(1-f_{L}(\varepsilon_{\uparrow}))+\Gamma_{R}(1-f_{R}(\varepsilon_{\uparrow}))$,
$T_{13}=\Gamma_{L}(1-f_{L}(\varepsilon_{\downarrow}))+\Gamma_{R}(1-f_{R}(\varepsilon_{\downarrow}))$,
$T_{14}=0$,
$T_{21}=\Gamma_{L}f_{L}(\varepsilon_{\uparrow})+\Gamma_{R}f_{R}(\varepsilon_{\uparrow})$,
$T_{22}=-\Gamma_{L}(1-f_{L}(\varepsilon_{\uparrow})+f_{L}(\varepsilon_{\downarrow}+U))-\Gamma_{R}(1-f_{R}(\varepsilon_{\uparrow})+f_{R}(\varepsilon_{\downarrow}+U))$,
$T_{23}=0$,
$T_{24}=\Gamma_{L}(1-f_{L}(\varepsilon_{\downarrow}+U))+\Gamma_{R}(1-f_{R}(\varepsilon_{\downarrow}+U))$,
$T_{31}=\Gamma_{L}f_{L}(\varepsilon_{\downarrow})+\Gamma_{R}f_{R}(\varepsilon_{\downarrow})$,
$T_{32}=0$, $T_{33}=-\Gamma_{L}(1+f_{L}(\varepsilon_{\uparrow}+U)-f_{L}(\varepsilon_{\downarrow}))-\Gamma_{R}(1+f_{R}(\varepsilon_{\uparrow}+U)-f_{R}(\varepsilon_{\downarrow}))$,
$T_{34}=\Gamma_{L}(1-f_{L}(\varepsilon_{\uparrow}+U))+\Gamma_{R}(1-f_{R}(\varepsilon_{\uparrow}+U))$,
$T_{41}=0$,
$T_{42}= \Gamma_{L}f_{L}(\varepsilon_{\downarrow}+U)+\Gamma_{R}f_{R}(\varepsilon_{\downarrow}+U)$,
$T_{43}=\Gamma_{L}f_{L}(\varepsilon_{\uparrow}+U)+\Gamma_{R}f_{R}(\varepsilon_{\uparrow}+U)$ and
$T_{44}=-\Gamma_{L}(2-f_{L}(\varepsilon_{\uparrow}+U)-f_{L}(\varepsilon_{\downarrow}+U))-\Gamma_{R}(2-f_{R}(\varepsilon_{\uparrow}+U)-f_{R}(\varepsilon_{\downarrow}+U))$.

\end{document}